\documentclass[prd,amsfonts,onecolumn,superscriptaddress,aps,nofootinbib,11pt]{revtex4-1}

\usepackage{amsmath, amsthm, amssymb, amsfonts, mathrsfs}
\usepackage{mathtools}
\usepackage{braket, slashed, bm}
\usepackage{hyperref}
\usepackage{booktabs}
\usepackage{scalefnt}
\usepackage{setspace}
\usepackage{amsmath, amsthm, amssymb, amsfonts, mathrsfs}
\usepackage{scalefnt}
\usepackage{braket, slashed, bm}
\usepackage{bbold}
\usepackage{listings}
\usepackage{siunitx}
\usepackage{booktabs}
\usepackage{soul}
\usepackage{color}
\usepackage[dvipsnames]{xcolor}
\usepackage{graphicx}
\usepackage{float}
\usepackage{placeins}
\usepackage{booktabs}
\usepackage{hyperref}
\usepackage[caption=false]{subfig}
\usepackage{soul} 
\usepackage{ulem}

\hypersetup{
	unicode=false,          
	pdftoolbar=true,        
	pdfmenubar=true,        
	pdffitwindow=false,     
	pdfauthor={William},     
	colorlinks=true,       
	linkcolor=cyan,          
	citecolor=magenta,        
	urlcolor=blue
}

\begin{document}

\title{Observing Double Higgs Production at the LHC \\ via Neutrino Feature Engineering in $hh\to b\bar{b}\ell\ell\nu\nu$ }

\author{Alexandre Alves}
\email{aalves@unifesp.br}
\affiliation{Departamento de F\'isica, Universidade Federal de S\~ao Paulo, UNIFESP, 09972-270, Diadema-SP, Brazil}
\affiliation{Centro de Ci\^encias Naturais e Humanas, Universidade Federal do ABC,\\
UFABC, 09210-580, Santo Andr\'e-SP, Brazil}
\author{Eduardo da Silva Almeida}
\email{almeidae@ufba.br}
\affiliation{Departamento de Física do Estado Sólido, Universidade Federal da Bahia,\\
UFBA, 40170-115, Salvador-BA, Brazil }
\author{Diego S. V. Gon\c{c}alves}
\email{diego.vieira@ufabc.edu.br}
\affiliation{Centro de Ci\^encias Naturais e Humanas, Universidade Federal do ABC,\\
UFABC, 09210-580, Santo Andr\'e-SP, Brazil}

\date{\today}
	
\begin{abstract}
    Double Higgs production is challenging even at the High Luminosity LHC. The Standard Model (SM) $hh\to b\bar{b}WW(ZZ)\to b\bar{b}\ell\ell\nu\nu$ has a moderate cross section compared to other decay modes, but its backgrounds, mainly top quark pairs, are overwhelming. In this work, we propose new kinematic features designed to improve the discrimination of double Higgs pairs, in the $b\bar{b}\ell\ell\nu\nu$ channel, with cut-based and multivariate analyses. The new features are built with the neutrinos' momenta solutions obtained from imposing mass constraints when calculating {\it Higgsness} and {\it Topness} variables. For the SM $hh$ production, we estimate a $3.7\sigma$ statistical significance from an optimized cut-based strategy, improving about 20\% over the best estimates of the literature, and $5\sigma$ from a multivariate analysis if systematic uncertainties on the backgrounds are small. The new variables are constructed as ratios of kinematic functions of the particles' momenta, being less prone to systematic errors. We also demonstrate the usefulness of the solutions in reconstructing heavy scalar resonances and other variables of phenomenological importance. 
\end{abstract}

\maketitle 
	
\section{Introduction}

 Discovering a new particle at colliders, like the Higgs boson back in 2012~\cite{ATLAS:2012yve, CMS:2012qbp}, is just the first step in the endeavor of truly knowing the new phenomenon. Properties related to that particle need to be established, such as its mass, quantum numbers including spin and parity, couplings to other particles of the spectrum, and, possibly, self-couplings. In the case of the SM Higgs, the LHC has already collected data to measure almost all those properties, except for couplings to light particles, such as electrons, up, and down quarks~\cite{Cepeda:2019klc}. 
 
 Another vital piece of information about Higgs bosons is still lacking: their self-interactions. In the SM, these self-interactions come from the scalar sector whose interaction Lagrangian after electroweak symmetry breaking reads
 \begin{equation}
{\cal
L}_h = \frac{1}{2}m_h^2 h^2+\lambda_{SM} v h^3+\frac{1}{4}\lambda_{SM} h^4\; ,
\label{eq:lagrangian}
 \end{equation}
 where $\lambda_{SM} = m_h^2/(2v^2)$, and $v\approx 246$ GeV is the Higgs vacuum expectation value, and $h$ is the CP-even scalar representing the Higgs field. Currently, combining search channels provides the best 95\% Confidence Level (CL) constraints on the modifier of the trilinear coupling, $\kappa_3=\lambda/\lambda_{SM}$, to be $-1.2<\kappa_3<7.2$ from ATLAS~\cite{CMS:2024awa, ATLAS:2024ish, ATLAS:2024xcs}; and  $-1.4<\kappa_3<7.8$ from CMS~\cite{CMS:2024awa, ATLAS:2024ish, ATLAS:2024xcs}, both from $\sim$ 140 fb$^{-1}$ of data collected from the 13 TeV LHC.

 The most straightforward way to observe the triple interaction and, in this way, to probe the $\lambda$ parameter, is by pair production of Higgs bosons. There are two contributions: one from the trilinear interaction and another from a top quark box. When $\lambda$ is positive, the interference between the amplitudes is destructive, and by a whim of nature, when the coupling is of the SM size, the destructive interference almost reaches its maximum, suppressing the cross-section. The NNLO QCD production cross section for $hh$ at the 14 TeV LHC is 40.7 fb~\cite{Grigo:2014jma}, which amounts to $\sim 10^5$ events produced at the LHC after 3000 fb$^{-1}$. The decay of the Higgs bosons dilutes that cross section, and the task now is finding a channel that, at the same time, is not suppressed by too rare a decay and is also not buried beneath huge backgrounds. The best options investigated in the literature are $hh\to b\bar{b}b\bar{b}$~\cite{CMS:2022cpr, ATLAS:2023qzf},  $hh\to b\bar{b}\gamma\gamma$~\cite{CMS:2020tkr, ATLAS:2025hhd},  $hh\to b\bar{b}\tau\tau$~\cite{CMS:2022hgz, ATLAS:2022xzm}, and  $hh\to b\bar{b}WW$~\cite{Baglio:2012np, CMS:2017rpp, Adhikary:2017jtu, Kim:2018cxf, Huang:2017jws, Huang:2022rne, Abdughani:2020xfo, Kim:2019wns}. The latter option has the second-best branching ratio after the four $b$-jets channel, but the $W\to\ell\nu$ decays suppress the total cross section to ${\cal O}(10^{-2})$. Taking all the branching ratios into account, the $hh\to b\bar{b}WW(ZZ)\to b\bar{b}\ell\ell\nu\nu$ cross section is 0.42 fb or around 1000 events at the end of the experiment. We also included subdominant decays $ZZ$, but as we will see, kinematic cuts will suppress them; therefore, for all practical purposes, only the channel $WW$ is important.

 Despite its reduced number of events, the authors of Ref.~\cite{Kim:2018cxf} showed that it is possible to reject the majority of the backgrounds for the $hh\to b\bar{b}WW$ channel, mainly the top quark pair and Drell-Yan production in the leptonic channel, by constructing two new variables called {\it Higgsness} and {\it Topness}, which measure how close the events are to their characteristics mass constraints\footnote{The ATLAS Collaboration has a public note showing plots for these variables in the semileptonic channel~\cite{ATLAS:2019mwn}.}. For example, in double Higgs events $M(WW)=M(\ell\ell\nu\nu)\sim m_h$, while for top pairs, $M(b\ell\nu)\sim m_t$. In both types of events, also $M(\ell\nu)\sim m_W$.
 
 In the process of computing the compatibility of the event with a given set of mass constraints, the neutrinos' momenta are obtained via minimization of {\it Higgsness} and {\it Topness}. Our work proposes exploring these solutions to construct variables that further separate the event classes. Not only can new variables be built, but key kinematic functions that are unavailable when neutrinos are lost can also be utilized. For example, if a new Higgs boson is present in the data, its invariant mass peak would be lost even though many other variables are at our disposal in this case~\cite{Franceschini:2022vck}.

 The variables that we are going to construct are physically and experimentally motivated. For example, invariant masses involving neutrinos' momentum and angular variables built in recovered frames of reference. Their combination follows from physics insight and inspection, although genetic programming and symbolic regression could be used to create better combinations, for example. We did not pursue that possibility and left it for future investigations. 

 The main result of our work is that the new variables can increase the prospects of finding $pp\to hh\to bbWW$ in the pure leptonic mode. Compared to Ref.~\cite{Kim:2018cxf}, our cut-based analysis improves the signal significance by $\sim$ 20\%, reaching $3.7\sigma$. In a profile-likelihood analysis with BDT scores, the significance surpasses the discovery threshold with no systematics in the background yields. Keeping these systematics at the 5\% level, we estimate that $hh$ can be discovered in this channel alone after 3000 fb$^{-1}$. Finally, we show the usefulness of the solutions in reconstructing the resonance of a new Higgs boson decaying to $hh$ and the cosine of the angle between the bottom jets' 3-momentum in the reference frame where the Higgs bosons are back-to-back.

 The paper is organized as follows: in Section~\ref{sec:2}, we describe our simulations; Section~\ref{sec:3} is devoted to the feature engineering of kinematic variables involving reconstructed neutrinos; in Section~\ref{sec:4} we present the results of the cut-based analysis, while Section~\ref{sec:5} presents the results of the multivariate analysis; the usefulness of the solutions beyond $hh$ discovery is illustrated in Section~\ref{sec:6}; we present our conclusions in Section~\ref{sec:7}.

 \section{Signal and Background Simulation}
 \label{sec:2}

  We simulate $pp\to hh\to b\bar{b}WW(ZZ)\to b\bar{b}\ell\ell\nu\nu,\; \ell=e,\mu,\tau$, and all the backgrounds with \texttt{MadGraph5}~\cite{Alwall:2014hca} at Leading Order QCD. Hadronization of jets is simulated with \texttt{Pythia8}~\cite{Sjostrand:2007gs} while jet clustering is performed with \texttt{FastJet}~\cite{Cacciari:2011ma} in the anti-$k_t$ mode with a fixed size jet cone of $R=0.4$. Detector effects are taken into account with \texttt{Delphes3}~\cite{deFavereau:2013fsa} using the default card. 

  The background sources we consider are the same as Ref.~\cite{Kim:2018cxf}, whose results represent our benchmark point. They are the following in decreasing importance order after selection strategies: $t\bar{t}$ \footnote{The dominant background is $t\bar{t}$; so, we simulate about 50 million events to obtain a reliable estimate. For $t\bar{t}h$, we simulate 2.5 million. For $\tau\tau\bar{b}b$, we generate 4 million events, while for $t\bar{t}Z$ and $t\bar{t}W$ the samples consist of about 2 million each. The DY sample contains 7 million events.}, $t\bar{t}h$, $t\bar{t}W$, $t\bar{t}Z$, and the Drell-Yan processes $\ell\ell+b(j)\bar{b}$ and $\tau\tau b\bar{b}$ where $\ell=e,\mu$. They also consider the mixed QCD and EW corrections to the Drell-Yan process. We did not simulate those events but just added their cross section, which is the smallest among the backgrounds, to the $\ell\ell bj$ background.
  The production cross sections, including relevant branching ratios into leptons $\ell=e,\mu,\; \hbox{and}\; \tau\to \nu\nu\ell$, and bottom quark pairs, are given in the first row of Table~\ref{tabela:xsecs}.

\begin{table}[t]
\centering
\resizebox{\textwidth}{!}{%
\begin{tabular}{ccccccc}
\hline\hline
$hh$ & $t\bar{t}$ & $t\bar{t}h$ & $ttW$ & $t\bar{t}Z$ & $\ell\ell b\bar{b}(jj)$ & $\tau^+\tau^- b\bar{b}$ \\
\hline\hline
 0.648 & $63.8\times 10^3$ & $23.7$ & $12.7$ & $27.7$ & $1.65\times 10^6$ & $2.03\times 10^4$ \\
 0.012 & 1.172 & 0.030 & 0.015 & 0.010 & 0.022 & 0.038 \\ 
\hline\hline
\end{tabular}%
}
\caption{In the upper row, we present the production cross sections in fb after decay into leptons, neutrinos, and bottom quarks, while in the lower row, the cross sections after cuts of Eq.~\eqref{eq:matchev_cuts}. We used the latter cross sections to normalize ours in order to compare the results of this study to those of Ref.~\cite{Kim:2018cxf}.}
\label{tabela:xsecs}
\end{table}

  We found a rough agreement between our results and those of Ref.~\cite{Kim:2018cxf} after imposing their selection requirements
  \begin{eqnarray}
&& \slashed{E}_T>20\; \text{GeV},\; p^{\ell}_T>20\; \text{GeV}\nonumber \\
&& \Delta R(\ell\ell)<1.0, \; m_{\ell\ell} < 65\; \text{GeV}\nonumber \\
&& \Delta R(bb) < 1.3, \; 95 < m_{bb} < 140\; \text{GeV.}
\label{eq:matchev_cuts}
  \end{eqnarray}
 We display, in the second row of Table~\ref{tabela:xsecs}, the cross sections after cuts of Eq.~\eqref{eq:matchev_cuts}. Because we want to compare our results to those of Ref.~\cite{Kim:2018cxf}, we normalize our cross sections by Table~\ref{tabela:xsecs}.

 \section{Feature Engineering with Neutrinos Momenta Solutions}
 \label{sec:3}

 \subsection{Neutrino solutions from {\it Higgsness} and {\it Topness}}

 The first step in constructing our variables is recovering some information from missing neutrinos. When two neutrinos are missing, six equations are needed for reconstruction, in addition to the neutrinos' mass constraints $p_\nu^2=0$ and $p_{\bar{\nu}}^2=0$. For $hh$ events, the neutrinos and leptons should reconstruct $W$ bosons, and $W^+W^-$ should reconstruct a Higgs boson, which provides four equations. Moreover, at least one of the $W$ bosons is off its mass shell. For top quarks, a neutrino+lepton should reconstruct a $W$ boson, which, in this case, is in its mass shell, and $W^+b$ reconstructs a top quark. Whatever the case, the components of the two neutrinos' momenta cannot all be recovered.

 An approximate solution can be obtained, though indirectly, when calculating the {\it Higgsness} and {\it Topness} variables~\cite{Kim:2018cxf}. {\it Higgsness} is defined as 
 \begin{eqnarray}
     H &\equiv& \underset{\vec{p}_\nu, \vec{p}_{\bar{\nu}}}{\mathrm{argmin}} \left[\frac{(M_{\ell^+\ell^-\nu\bar{\nu}}^2 - m_h^2)^2}{\sigma_h^4} +
     \frac{(M^2_{\nu\bar{\nu}}-M^2_{\nu\bar{\nu},peak})^2}{\sigma_\nu^4} \right.\nonumber\\
     &+& \left. \min\left(\frac{(M^2_{\ell^+\nu}-m_W^2)^2}{\sigma_W^4}+\frac{(M^2_{\ell^-\bar{\nu}}-m_{W^*,peak}^2)^2}{\sigma_{W^*}^4},\; \frac{(M^2_{\ell^-\bar{\nu}}-m_W^2)^2}{\sigma_W^4}+\frac{(M^2_{\ell^+\nu}-m_{W^*,peak}^2)^2}{\sigma_{W^*}^4}\right)\right], \nonumber \\
     &&
     \label{eq:higgsness}
 \end{eqnarray}
 where $\sigma_h$, $\sigma_W$, and $\sigma_\nu$ might represent experimental uncertainties (in GeV), but for our purposes, they can be treated as free parameters. 
 The peaks of the $M_{\nu\bar{\nu}}$ and $M_{W^*}$ distributions occur approximately at 37 and 31 GeV, respectively. We fixed $\sigma_h=2$ GeV, $\sigma_W=\sigma_{W^*}=5$ GeV, and $\sigma_\nu=10$ GeV as in Ref.~\cite{Kim:2018cxf}.

 {\it Topness}, in its turn, is given by
 \begin{eqnarray}
     T & \equiv & \min(\chi^2_{12},\chi^2_{21}) \nonumber \\
     \chi^2_{ij} &=& \underset{\vec{p}_\nu, \vec{p}_{\bar{\nu}}}{\mathrm{argmin}}\left[\frac{(M^2_{b_i\ell^+\nu}-m^2_t)^2}{\sigma_t^4}+\frac{(M^2_{\ell^+\nu}-m^2_W)^2}{\sigma_W^4}+\frac{(M^2_{b_j\ell^-\bar{\nu}}-m^2_t)^2}{\sigma_t^4}+\frac{(M^2_{\ell^-\bar{\nu}}-m^2_W)^2}{\sigma_W^4}\right]\; ,
     \label{eq:topness}
 \end{eqnarray}
 where $\sigma_t=5$ GeV. In this case, as we do not know the $b$-jet charge, we have to test between two options to get the better consistency of the event with the $t\bar{t}$ production and decay chain $t(\bar{t})\to W^+ b(W^-\bar{b})\to b\ell^+\nu(\bar{b}\ell^-\bar{\nu})$.

 The minimization is carried out on the neutrinos' 3-momenta while their energies are determined from $E=|\vec{p}|$. It is possible to use the missing transverse momenta to constrain the solutions even further, but we chose to keep all six unknowns from $\vec{p}_\nu, \vec{p}_{\bar{\nu}}$ in $H$ and $T$. We used the simplex algorithm~\cite{nelder1965simplex} from the \texttt{Scipy} implementation~\cite{2020SciPy-NMeth} for minimization.

\begin{figure}[t]
    \centering
    \includegraphics[width=1.0\linewidth]{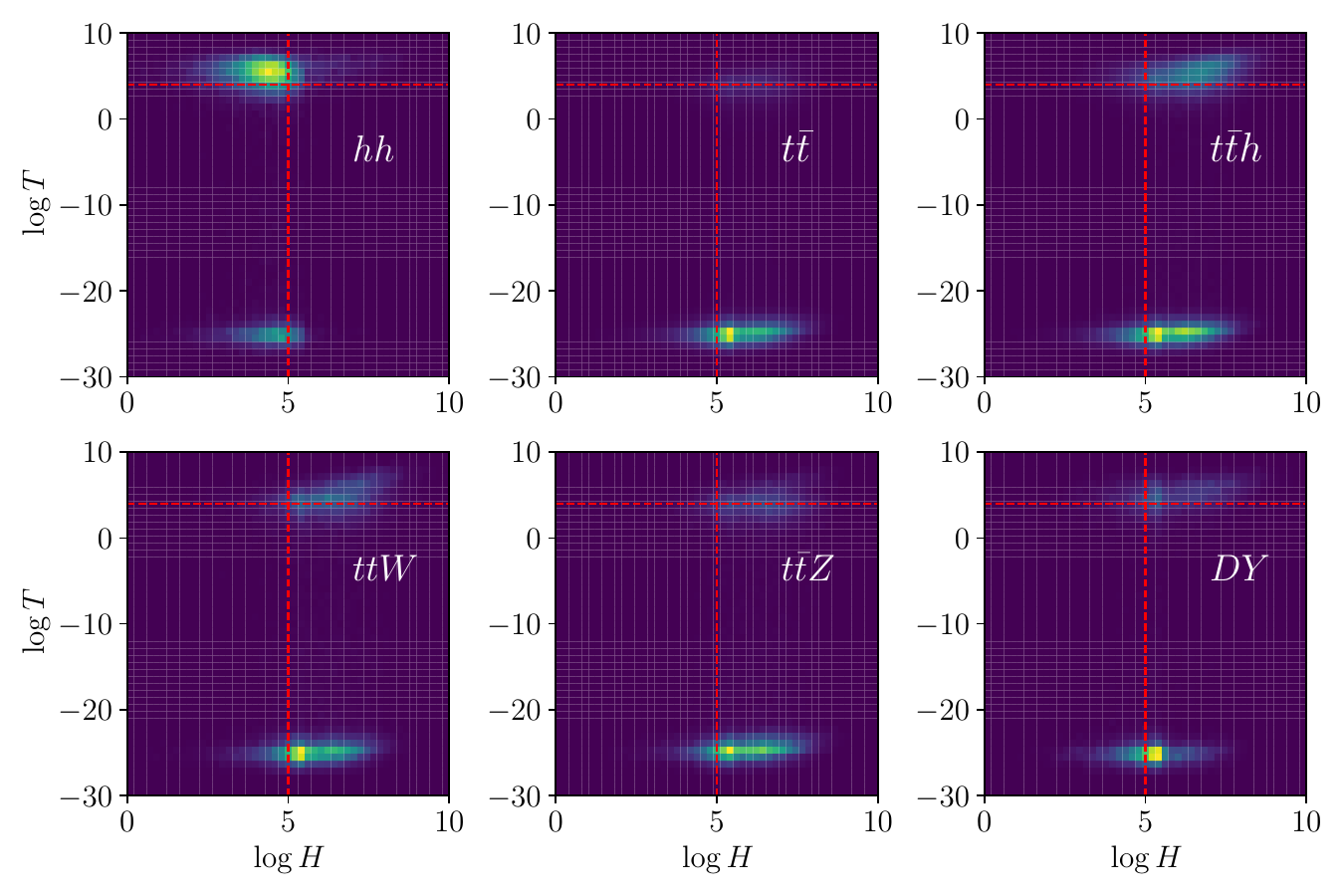} 
\caption{The density of events in the log of {\it Higgsness} versus the log {\it Topness} plane for all class events considered in this work. The red dashed lines pass through $\log H=5$ and $\log T=5$ and help to visualize the regions in the plane where the events cluster.}
    \label{fig:HT}
\end{figure}

 In Figure~\ref{fig:HT}, we show the distributions of the logarithm of {\it Higgsness} and {\it Topness}. While $hh$ events are concentrated on the upper-left corner of the $\log H\times\log T$ plane, the majority of backgrounds are very top-like with negative values of {\it Topness}. Notice, however, that a small mass of $t\bar{t}$ and Drell-Yan events, compared to the region of higher concentration in the upper central panel, lies in the signal region. The number of events from these processes is huge, so even this small contamination has to be tamed for us to have a chance to observe the Higgs pair production in this channel. 

\subsection{Kinematic Variables Constructed with Neutrino Solutions}

 Now that we have two sets of neutrino solutions obtained from Eqs.~\eqref{eq:higgsness} and ~\eqref{eq:topness}, we can construct kinematic variables that might be useful in distinguishing signal and background events. There are many possibilities, for example, by calculating the mass of particle systems that include neutrinos as the center-of-mass energy of the incoming partons, $\sqrt{\hat{s}}$, the top mass, and $h\to WW$ mass. However, there are two options for all of them: constructions that involve the {\it Higgsness} solution $\nu_h$ or the {\it Topness} solution $\nu_t$. We can also select solutions based on $\log H$ and $\log T$ of the event, for example, $\nu=\nu_h$ if $(\log H,\log T)\in HR$, and $\nu=\nu_t$ if $(\log H,\log T)\in TR$, where $HR$($TR$) is a given Higgs(Top) region of the $\log H\times\log T$ plane.

\begin{figure}[t]
    \centering
    \includegraphics[width=1.1\linewidth]{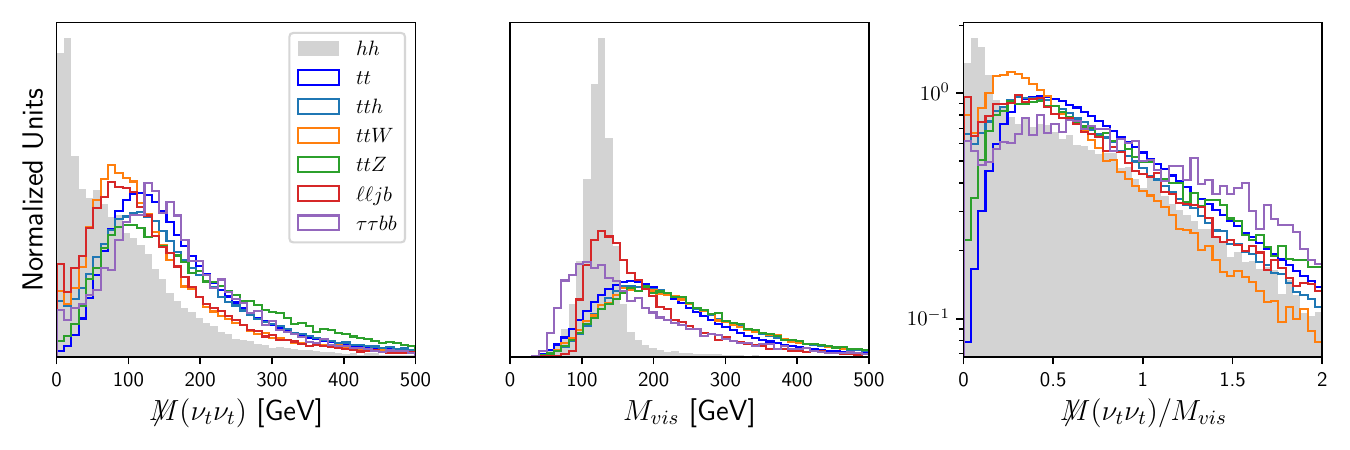}\\
    \includegraphics[width=1.1\linewidth]{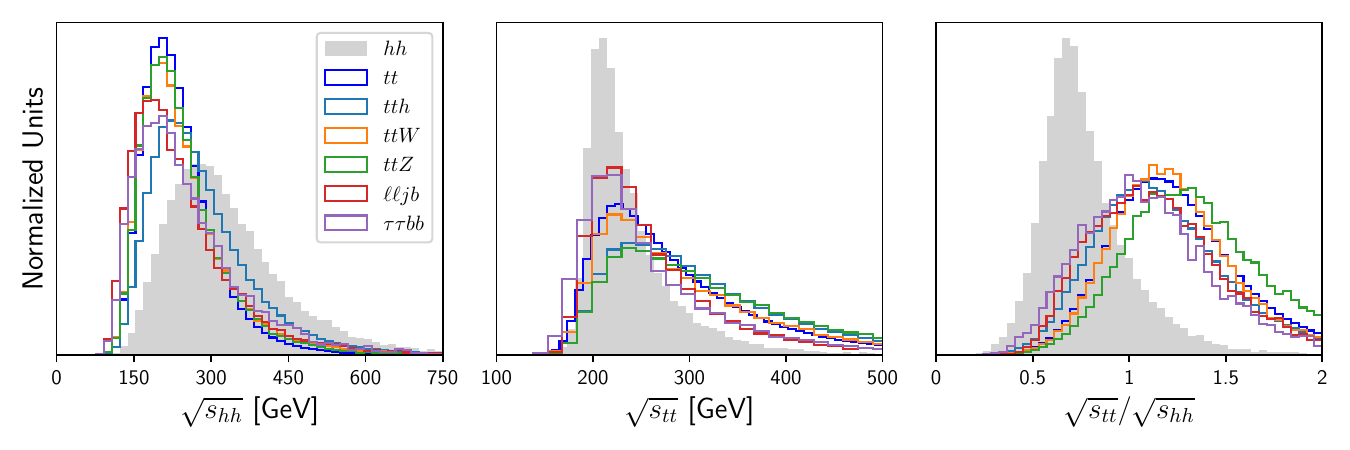}\\
    \includegraphics[width=1.1\linewidth]{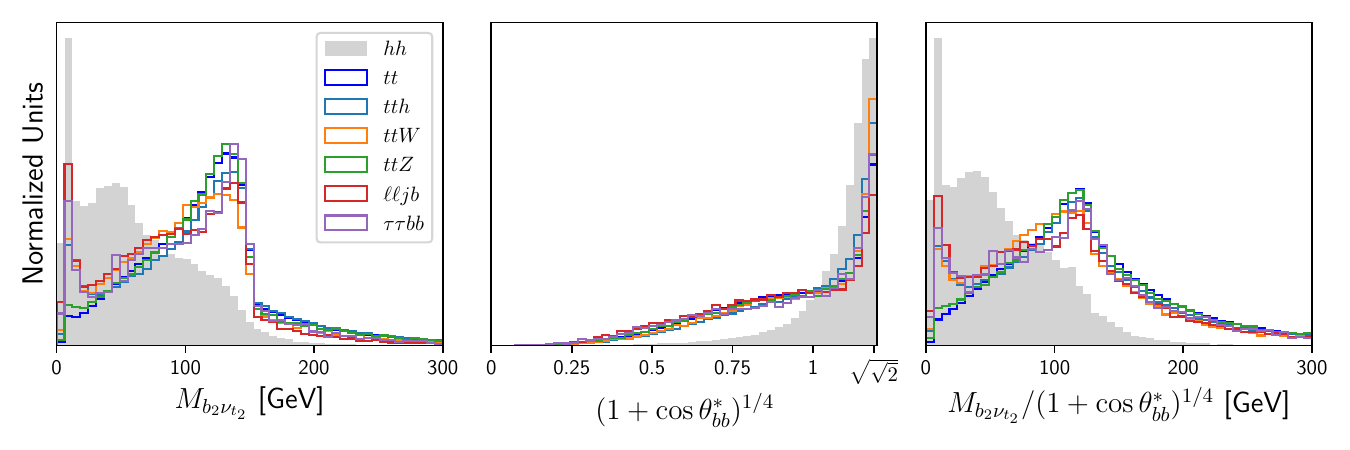}
\caption{The three new variables we propose can be seen in the rightmost panels of this plot. The other panels display the variables from which the new ones are constructed.}
    \label{fig:NewVariables}
\end{figure}

 We now present some variables that we found very effective in separating signals from backgrounds, especially from the top-quark pair ones. In Figure~\ref{fig:NewVariables}, we display three new variables that involve neutrino momentum solutions. In the left panel, we have the invariant mass of the two {\it Topness} neutrino solutions in the first row $\slashed{M}(\nu_t\nu_t)$. The salient characteristic of this variable is the sharp peak of the $hh$ events at $\slashed{M} (\nu_t\nu_t)\approx 0$. This peak is due to solutions where $p_\nu,p_{\bar{\nu}}\approx 0$ for events that do not satisfy the top quark pairs constraints, mainly the $hh$ events. We provide a more detailed explanation in Appendix~\ref{apendiceA}. In the center panel of the first row, we see the visible mass of the event, defined as $M_{vis}=\sqrt{M_{\ell\ell}^2+M_{bb}^2}$. The new variable, let us call it $RIV_t$, is given by the ratio 
 \begin{equation}
 RIV_{t} = \dfrac{\slashed{M}(\nu_t\nu_t)}{M_{vis}},
 \label{eq:RIVt}
 \end{equation} 
 and is displayed on the rightmost panel of the first row. The ratio enhances the difference between the $hh$ and $t\bar{t}$ events in the low value bins. Another reason to choose a ratio in the feature engineering is that some systematic errors cancel in ratios, for example, as luminosity uncertainties, as long as the numerator and denominator are independent variables. We checked that the three new variables have low Pearson and Spearman $R$ coefficients (less than 0.1).

 In the second row of Figure~\ref{fig:NewVariables}, we have the $\sqrt{s_{hh}}$ and $\sqrt{s_{tt}}$ variables in the first and second columns, respectively. They are defined as
 \begin{eqnarray}
&& s_{hh} = (p_{\ell\ell}+p_{\nu\nu})\cdot p_{bb},\; s_{tt} = p_{\ell_1\nu_1 b_1}\cdot p_{\ell_2\nu_2 b_2}\label{eq:sht}\\
&& R_{ss} = \dfrac{\sqrt{s_{hh}}}{\sqrt{s_{tt}}}
 \end{eqnarray}

While the spectrum of $\sqrt{s_{hh}}$ is harder for $hh$ compared to backgrounds, the opposite is true concerning $\sqrt{s_{tt}}$. Therefore, for $hh$ events, the ratio $\sqrt{s_{tt}}/\sqrt{s_{hh}}$ tends to concentrate for low value bins, peaking around 0.75, and the backgrounds peak at 1.25, approximately. Again, the ratio guarantees a lower sensitivity to systematic uncertainties.

We show, in the lower row of Figure~\ref{fig:NewVariables}, the last of the three new variables that we will dig for $hh$ events, cleaning it from backgrounds, especially the top-quark ones. In the leftmost panel, the $M_{b_2\nu_{t_2}}$ invariant mass displays a sharp peak for $hh$ towards small value bins but peaks beyond 100 GeV for backgrounds. To enhance the peak, we divide it by $(1+\cos\theta^*_{bb})^{1/4}$, where
\begin{equation}
    \cos\theta^*_{bb} = \dfrac{\vec{p^*}_{b_1}\cdot\vec{p^*}_{b_2}}{|\vec{p^*}_{b_1}|\; |\vec{p^*}_{b_2}|},
    \label{eq:cost}
\end{equation}
where $\vec{p^*}$ represents a 3-momentum in the rest frame of the Higgs boson pair. This new variable is thus given by
\begin{equation}
    cM_{b_2\nu_{t_2}} = \dfrac{M_{b_2\nu_{t_2}}}{(1+\cos\theta^*_{bb})^{1/4}},
    \label{eq:cMbv}
\end{equation}
using the $\nu_{t_2}$ neutrino solution, the softest neutrino from {\it Topness}.

We clearly realize, looking at the rightmost panel of Figure~\ref{fig:NewVariables}, that, for backgrounds, the peak does not move much, but the tails become heavier compared to $M_{b_2\nu_{t_2}}$, that is, the events migrate to the tail, favoring a cut that keeps low values of the variable. This variable, in particular, was built by visually inspecting, but it demonstrates the potential for feature engineering using more advanced tools than the human eye, such as genetic programming and symbolic regression, for example.

We also show, in Appendix~\ref{apendiceB}, the quality in the reconstruction of the momentum components $(p_T,\eta,\phi)$ of the heavy particles of the events, the SM Higgs $h$, the $W$-boson, and the top quark $t$.
 
Now that we have constructed new distinctive variables, let us seek a cut-based strategy to separate signals from backgrounds.

\section{CUT-BASED ANALYSIS}
\label{sec:4}

 The basic selection is given by 
 \begin{eqnarray}
&& p_T(\ell)>20\hbox{ GeV},\; |\eta(\ell)|<2.5,\; p_T(b)>30\hbox{ GeV},\; |\eta(b)|<2.5,\nonumber\\
&& \slashed{E}_T>20\hbox{ GeV.}
\label{eq:basic-cuts}
 \end{eqnarray}

\begin{figure}[t]
    \centering
    \includegraphics[width=0.45\linewidth]{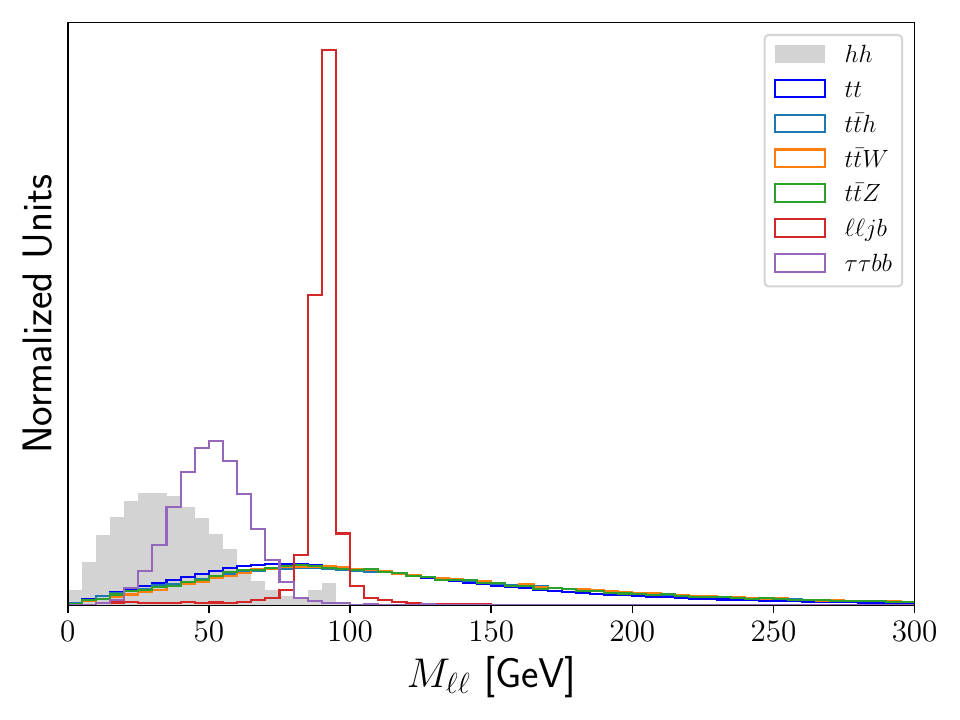}
    \includegraphics[width=0.45\linewidth]{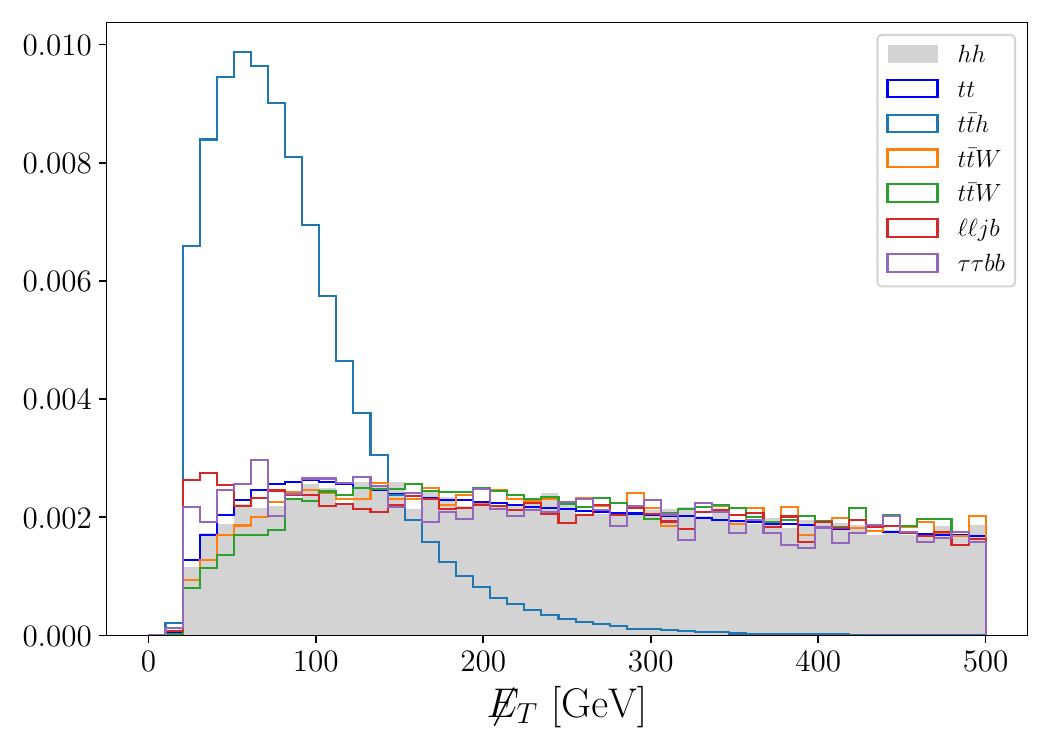}\\
    \includegraphics[width=0.45\linewidth]{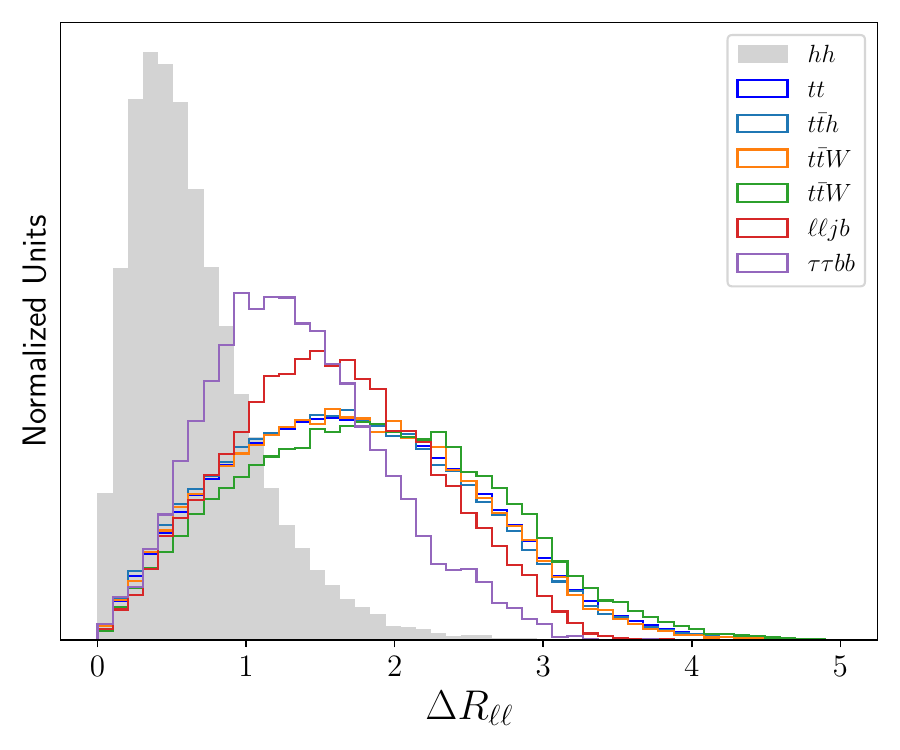}
    \includegraphics[width=0.45\linewidth]{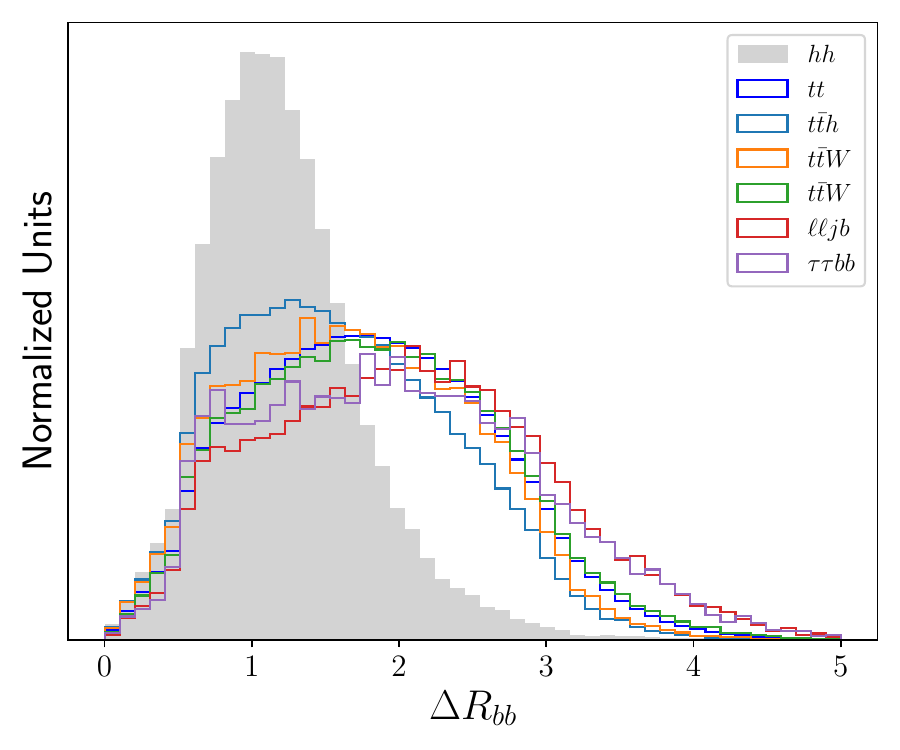}
\caption{Variables used in conjunction with those of Eqs.~\eqref{eq:RIVt}--\eqref{eq:cMbv} to perform the optmized cut-based analysis. In the upper row, we have the leptons invariant mass at left, and the missing transverse energy at right. In the lower panel, we have the distance in the $\eta-\phi$ plane between leptons at left, and bottom jets at the right.}
    \label{fig:OldVariables}
\end{figure}
 The optimization of cuts is performed on many combinations of variables. We used \texttt{Hyperopt}~\cite{hyperopt} to find the largest signal significance possible, calculated with the Asimov formula.
\[
Z_A(S,B,\sigma_B)
= \left\{
2\left[
(S+B)\,\ln\!\left(
\frac{(S+B)\,(B+\sigma_B^2)}{\,B^2+(S+B)\,\sigma_B^2\,}
\right)
- \frac{B^2}{\sigma_B^2}\,
\ln\!\left(
1+\frac{\sigma_B^2\,S}{B\,(B+\sigma_B^2)}
\right)
\right]
\right\}^{1/2},
\]
where $\sigma_B=\varepsilon_B B$ is the systematic error in the background yields.


The best set of variables comprised the three new variables proposed in this work, defined in Eqs.~\eqref{eq:RIVt}--\eqref{eq:cost} plus Higgness~\eqref{eq:higgsness}, Topness~\eqref{eq:topness}, and the kinematic variables $M_{\ell\ell}$, the invariant mass of the lepton pair, the distance, in the $\phi-\eta$ plane, $\Delta R_{\ell\ell}$, between the leptons, the distance, $\Delta R_{bb}$, between the bottom jets, and the missing transverse energy of the event $\slashed{E}_T$. These latter variables are displayed in Figure~\ref{fig:OldVariables}.

\begin{figure}[t]
    \centering
    \includegraphics[width=1.0\linewidth]{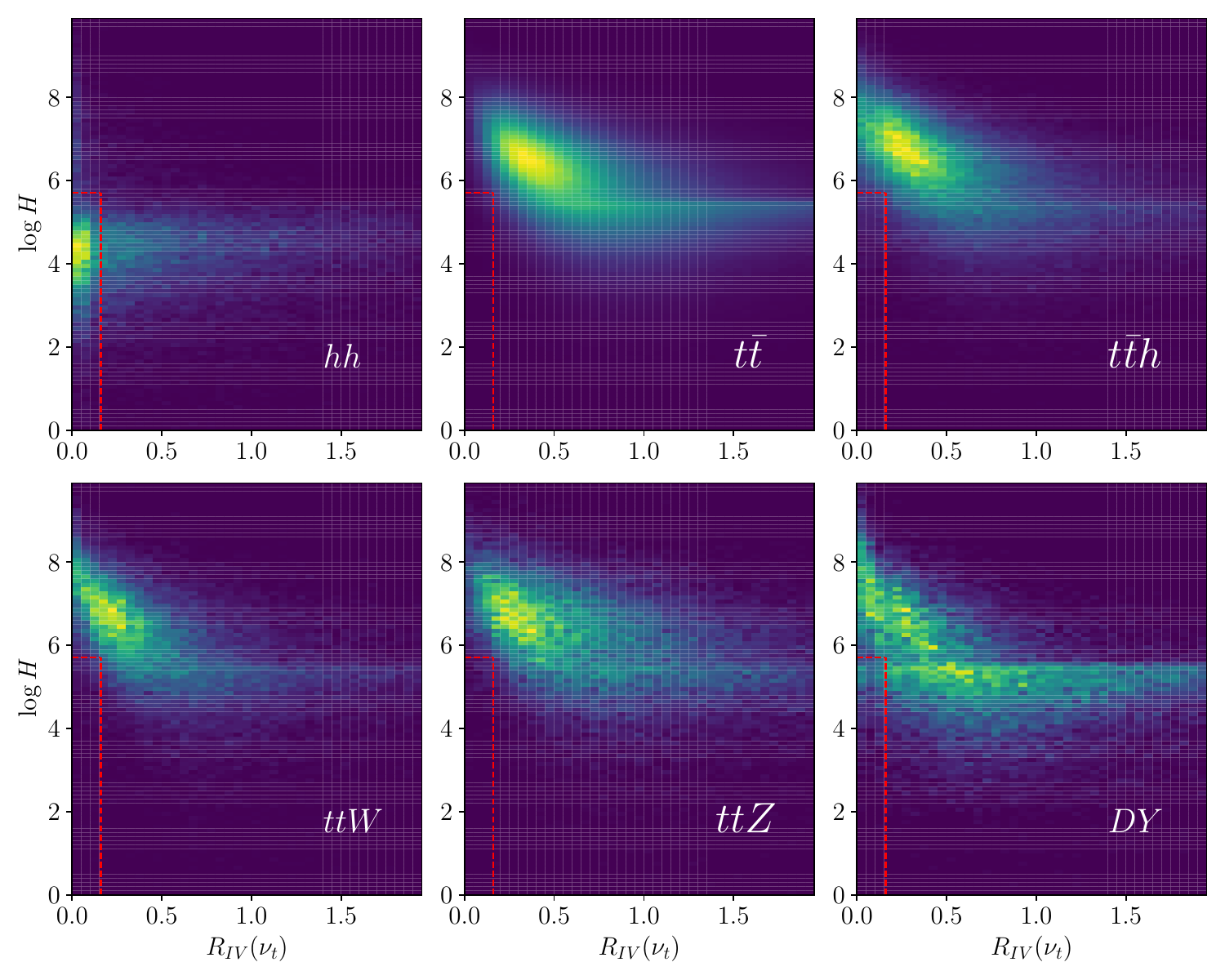}
\caption{The density of events in the $RIV_t$ versus $\log H$ plane for all event classes. The red dashed box indicates the region selected in the optimization of the signal significance.}
    \label{fig:RIVt-logH}
\end{figure}
The optimized cuts found are the following
\begin{eqnarray}
&&  \log T > 5.70,\; \log H < 4.83,\;  RIV_t < 0.16,\nonumber \\
&& cM_{b_2\nu_{t_2}} < 246.3\hbox{ GeV},\; M_{ll} < 53.5\hbox{ GeV},\;   \slashed{E}_T>87.9\hbox{ GeV,}\nonumber  \\
&& \Delta R_{\ell\ell}<0.32,\; \Delta R_{bb}<1.28,\; R_{ss} < 2.38\; ,
\label{eq:opt-cuts}
\end{eqnarray}
from which a statistical signal significance of $3.7\sigma$ is achieved with a signal-to-background ratio of 6.6, equivalent to 5 expected signal events and 0.75 background events.
A $3.6\sigma$ with $S/B\sim 2$, $S=10.3$ and $B=5.3$ events can be found with very similar cuts, but relaxing $\log T<4.60$. A somewhat degraded result of $\sim 3.5\sigma$, with high $S/B$ though, is also possible by removing the $R_{ss}$ variable. A $3\sigma$ significance with various $S/B$ can be reached without the variables presented here, as demonstrated in Ref.~\cite{Kim:2018cxf}. All these results apply after 3000 fb$^{-1}$.

The most effective variables to eliminate background events are $\log H$ and $RIV_{t}$, as we see in Figure~\ref{fig:RIVt-logH}. The signal events concentrate in the lower left corner of the plane, while the backgrounds, notably $t\bar{t}$, at the complementary region. In Table~\ref{tabela:cuts}, we show the number of events of each class for an integrated luminosity of 3000 fb$^{-1}$ after placing the optimized cuts sequentially.

\begin{table}[t!]
\centering
\resizebox{\textwidth}{!}{%
\begin{tabular}{cccccccc}
\hline\hline
cut & $hh$ & $t\bar{t}$ & $t\bar{t}h$ & $ttW$ & $t\bar{t}Z$ & $\ell\ell b\bar{b}(jj)$ & $\tau^+\tau^- b\bar{b}$ \\
\hline\hline
basic & 102 & $3.65\times 10^5$ & $5.85\times 10^3$ & $4.90\times 10^3$ & $4.10\times 10^3$ & $1.10\times 10^5$ & $4.84\times 10^3$ \\
$cM_{b_2\nu_2}$ & 101 & $3.25\times 10^5$ & $5.29\times 10^3$ & $4.54\times 10^3$ & $3.64\times 10^3$ & $1.01\times 10^5$ & $4.84\times 10^3$ \\
$M_{\ell\ell}$ & 82 & $5.83\times 10^4$ & 794 & 638 & 455 & 2313 & 2771 \\
$RIV_t$ & 22 & 573 & 34 & 33 & 11 & 156 & 223 \\
$\Delta R_{\ell\ell},\Delta R_{bb}$ & 7.7 & 14 & 2.8 & 1.2 & 0.7 & 0 & 1.5 \\
$\slashed{E}_T$ & 7.3 & 13.4 & 1 & 1 & 0.7 & 0 & 1.46 \\
$\log H$, $\log T$ & 5 & 0.5 & 0.2 & 0.12 & 0 & 0 & 0 \\
$R_{ss}$ & 5 & 0.5 & 0.2 & 0.05 & 0 & 0 & 0 \\
\hline\hline
\end{tabular}%
}
\caption{Cutflow on the number of the signal and background events after 3000 fb$^{-1}$. In the first row of the Table, we display the number of events after basic cuts of Eq.~\eqref{eq:basic-cuts}. The subsequent rows show the surviving number of events after placing the cuts of Eq.~\eqref{eq:opt-cuts} sequentially.}
\label{tabela:cuts}
\end{table}
\section{Multivariate Analysis}
\label{sec:5}

Beyond cut-based analysis, we conduct a multivariate study that incorporates all the variables described in the previous section. Our strategy is to train a boosted decision tree (BDT) algorithm to separate the signal and background in a multiclass task. Instead of simply placing cuts on the BDT's output score, $p(y=hh|x)$, we calculated the log-likelihood ratio between the $S$ and $S+B$ distributions to leverage the profiles of the distributions.

The dataset was split into a 20\% hold-out test set and an 80\% training/validation pool using stratified sampling to preserve class distributions. Approximately four million events were used to train, validate, and test the \texttt{XGBoost}~\cite{Chen:2016:XST:2939672.2939785} implementation of boosted decision trees (BDTs). The input features were $\log H$, $\log T$, $M_{\ell\ell}$, and the new variables $RIV_t$, $R_{ss}$, and $cM_{b_2\nu_{t_2}}$. Hyperparameter optimization was carried out with Optuna~\cite{akiba2019optunanextgenerationhyperparameteroptimization}, tuning \texttt{max\_depth}, \texttt{learning\_rate}, \texttt{subsample}, \texttt{colsample\_bytree}, and \texttt{min\_child\_weight} over 50 trials, each allowing up to 50 estimators (fewer when stopped early by pruning callbacks or by early stopping after 10 rounds). To mitigate overfitting and ensure stable performance during tuning, each Optuna trial was evaluated with stratified three-fold cross-validation on the training/validation pool, and the average validation score across folds was used as the tuning objective. Class imbalance was addressed using \texttt{XGBoost}’s native class-weighting scheme. The final model was trained on the complete training/validation pool with the optimized hyperparameters and class weights, and then evaluated on the independent test set. Performance was assessed using log loss, standard and balanced accuracy, and per-class metrics. During this final stage, up to 1000 estimators were allowed, with pruning callbacks and early stopping after 50 rounds.

\begin{figure}[t]
    \centering
    \includegraphics[width=0.45\linewidth]{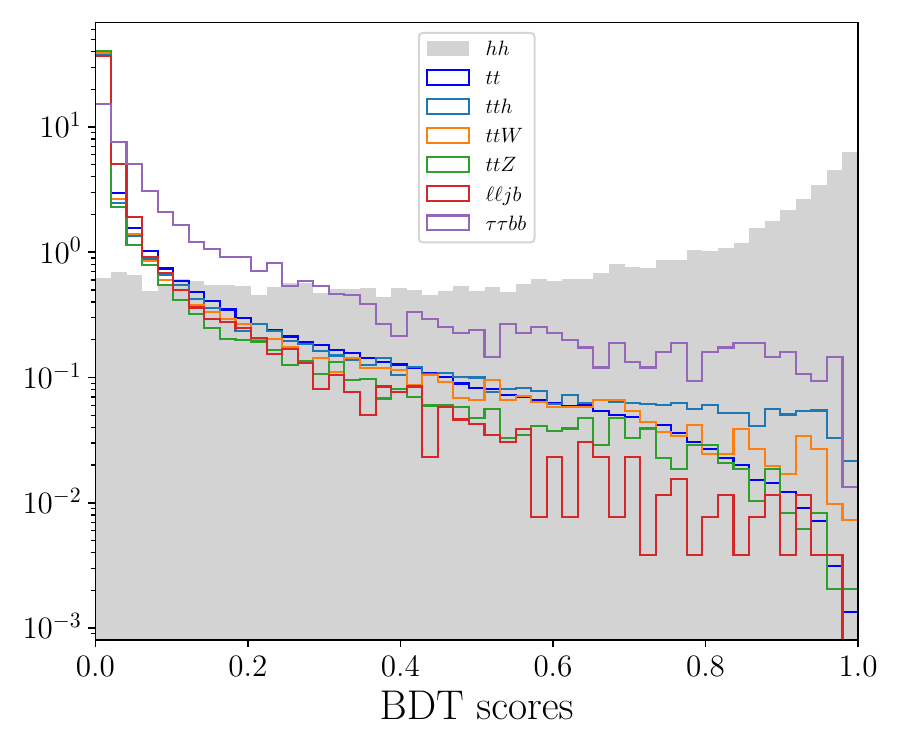}
    \includegraphics[width=0.45\linewidth]{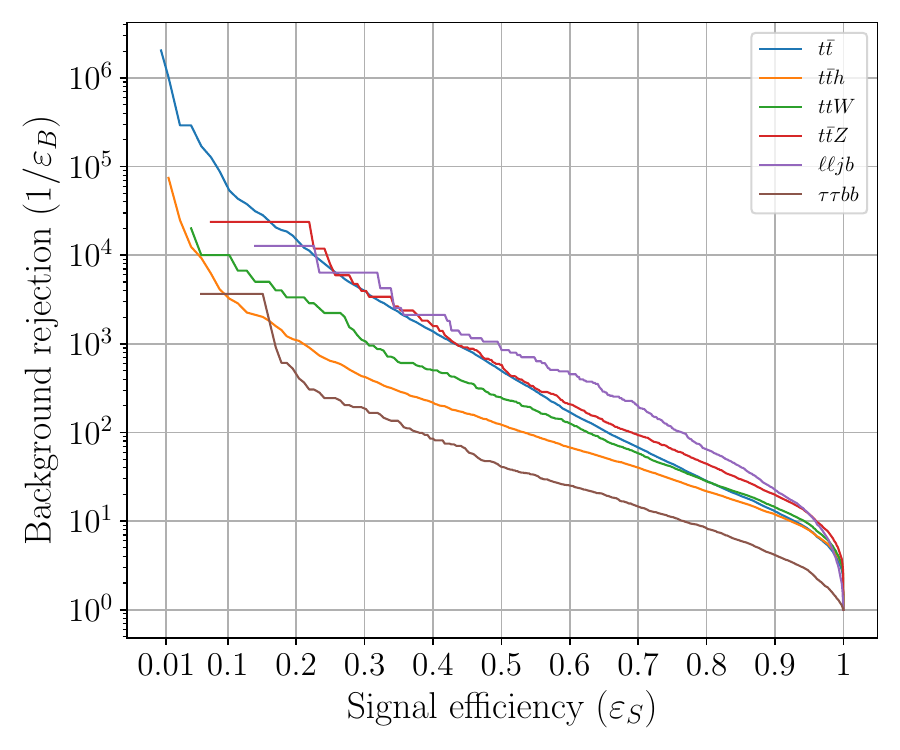}
\caption{BDT output scores, in the left panel, and background rejection versus signal efficiency curves in the right panel.}
    \label{fig:ML}
\end{figure}

In Figure~\ref{fig:ML}, in the left panel, we show the output score distribution of the $hh$ signal and the backgrounds. The $hh$ events peak strongly at $p(y=hh|x)\sim 1$, while all the other backgrounds, especially the most harmful, the top pair production, occur at $p(y=hh|x)\sim 0$. Also, backgrounds with a $Z$ boson are present due to characteristic peaks in the invariant mass of $\ell^+\ell^-$. In the right panel, we display the signal efficiency versus background rejection for all backgrounds. Two benchmark points show the exquisite job of the new variables allied to the BDT -- $\varepsilon_S(1/\varepsilon_B)=0.01(10^6)$, $\varepsilon_S(1/\varepsilon_B)=0.1(6\times 10^4)$, for $t\bar{t}$, and  $\varepsilon_S(1/\varepsilon_B)=0.01(8\times 10^4)$, $\varepsilon_S(1/\varepsilon_B)=0.1(3\times 10^3)$, for $t\bar{t}h$, the two dominant backgrounds after cuts of Eq.~\eqref{eq:basic-cuts}.

 It is possible to surpass the cut-based strategy after imposing moderate cuts on some key variables and then computing the profile likelihood statistics of the BDT output score with the $CL_{s}$ method using \texttt{pyhf}~\cite{pyhf_joss}. For example, with the cuts
 \begin{eqnarray}
     && M_{\ell\ell} < 80\hbox{ GeV},\; \log H<5,\; \log T>5, \\
     && RIV_t < 0.2,\; R_{ss}<0.9,
 \end{eqnarray}
 we reach $5\sigma$ with a systematic error of $\epsilon_{sys}=5$\% on the backgrounds, where the total uncertainty is given by $\delta B = \sqrt{B+(\epsilon_{sys} B)^2}$. This signal significance degrades rapidly as the background uncertainty increases, forcing us to tighten the cuts and make the distributions more similar. For $\epsilon_{sys}=1$\%, we have $N_\sigma\sim 7\sigma$, and for $\epsilon_{sys}=5$\%, we have $N_\sigma=5\sigma$, while if $\epsilon_{sys}=10$\%, we reach only $N_\sigma\sim 3.1\sigma$. So, it is important to keep the systematic uncertainties in the backgrounds under control to benefit from the shape analysis.

\begin{figure}
    \centering
    \includegraphics[width=0.4\linewidth]{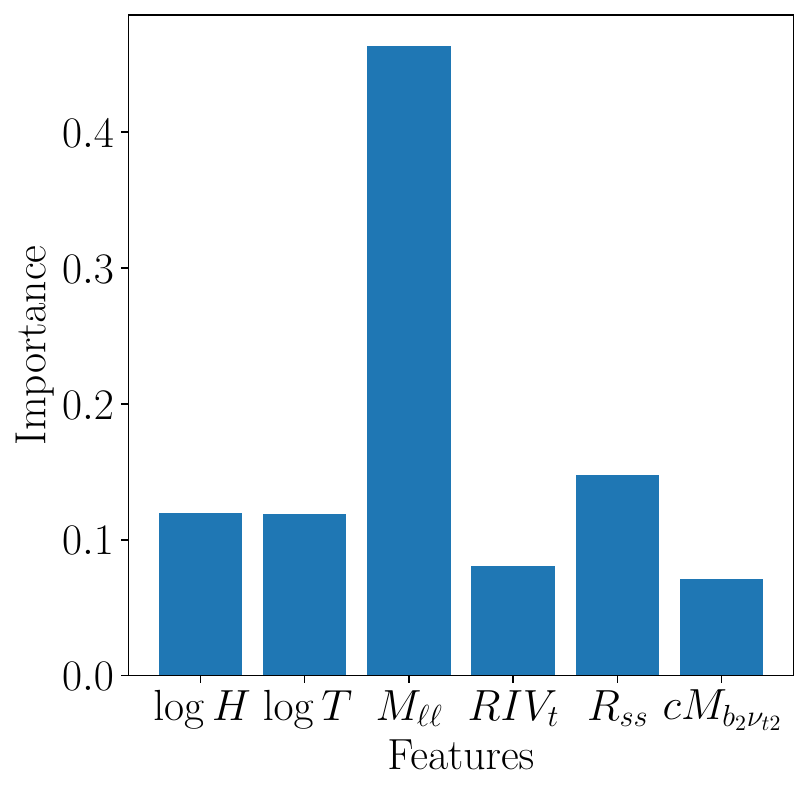}
    \includegraphics[width=0.465\linewidth]{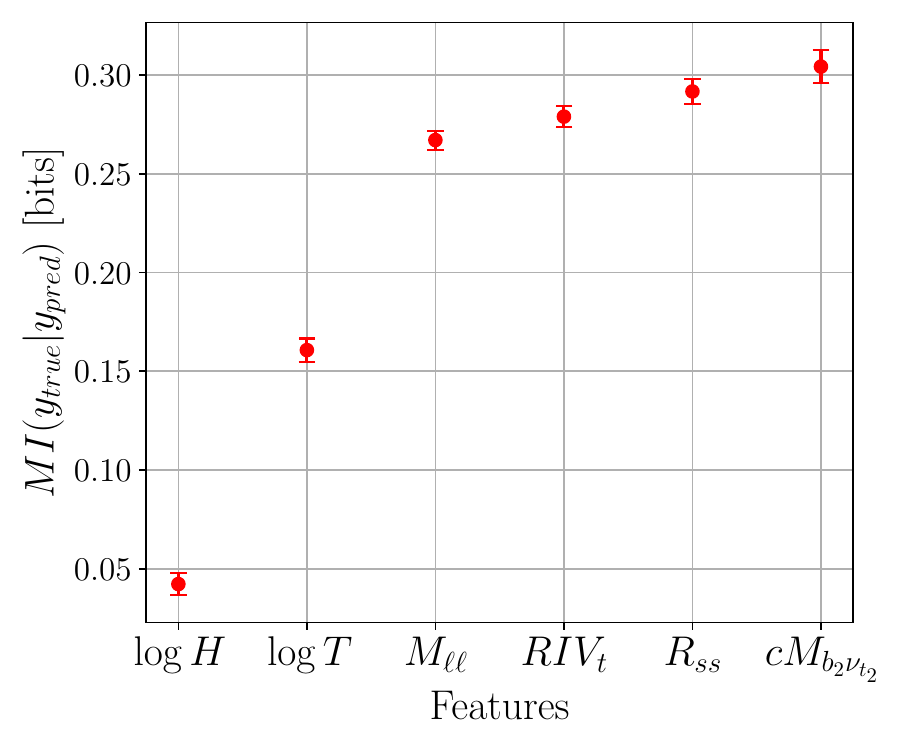}
    \caption{Left plot: feature importance classification of all variables used in the BDT with Shapley values. $M_{\ell \ell}$ stands out as the most important feature for its role in classifying the Drell-Yan events. The new variable $R_{ss}$ is the second most important feature. {\it Higgness} and {\it Topness} present equal importance. Right panel: the mutual information between the true and the predicted labels.}
    \label{fig:feat_importance}
\end{figure}

 We performed a feature importance analysis with Shapley values as implemented in \texttt{SHAP}~\cite{shap_python}, see Figure~\ref{fig:feat_importance}. The invariant mass of leptons has a prominent importance in distinguishing the events mainly due to the Drell-Yan resonance, as we see in Figure~\ref{fig:OldVariables}. The second more important variable is the new variable $R_{ss}$, followed by {\it Higgsness} and {\it Topness}, which, interestingly, have the same importance for the BDT classification. These last three variables have a more uniform power to discern amongst the classes.  Then we have $RIV_t$, and $cM_{b_2\nu_{t_2}}$ in the scale of importance. The $RIV_t$ and $cM_{b_2\nu_{t_2}}$ are important players to discard $t\bar{t}$ and $t\bar{t}h$ events by their turn.

 Another feature importance of great value is the computation of the mutual information between the BDT prediction, $y_{pred}$, and the true labels, $y_{true}$, of the events, $MI(y_{true};y_{pred})$. The mutual information is a nonlinear measure of correlation between the variables based on information theory, and it is given by
 \begin{equation}
     MI(y_{true};y_{pred}) = \sum_{y_{true}\in {\cal Y}}\sum_{y_{pred}\in {\cal Y}} p(y_{true},y_{pred})\log\dfrac{p(y_{true},y_{pred})}{p(y_{true})p(y_{pred})}\; ,
 \end{equation}
where ${\cal Y}$ is the support of the probability functions, $p(.)$.
 
 The mutual information is zero when $p(x,y)=p(x)p(y)$, that is, when two variables are independent from each other. If the BDT prediction, trained with a set of variables, is truly predictive of the true label of the events, the mutual information will be high as the correlation between them is strong. Adding more variables to the set might increase it, if it helps in the prediction, or decrease it, if not. This way,  $MI(y_{true};y_{pred})$ works as a model-free feature importance measure~\cite{CarraraErnst2017UpperLimitOfSeparability}. 

 We display, in Figure~\ref{fig:feat_importance}, lower panel, the mutual information of the BDT predictions and the true labels starting with the set of variables $\{\log H,\log T\}$, and adding, one at a time, the other variables.
 Confirming the result from Shapley values, adding $M_{\ell\ell}$ increases $MI$ more than the others, but the new ones keep increasing the mutual information, demonstrating that there's still information to be extracted to discern the classes despite the excellent job of {\it Higgsness}, {\it Topness}, and the invariant mass of the leptons. It is important to notice, though, that there is no direct relation between $MI$ and usual figures of merit (FOM) like the accuracy of the BDT predictions or the signal significance; yet it is expected that a higher mutual information corresponds to a better discriminant and improved FOM~\cite{CarraraErnst2017UpperLimitOfSeparability}. Finally, the data points in the lower panel of Figure~\ref{fig:feat_importance} vary as a feature of the algorithms used to approximately calculate $MI$~\cite{CarraraErnst2017UpperLimitOfSeparability}. To mitigate the variation, we calculate the mutual information 20 times and take its mean and standard deviation. Nevertheless, the means of the data points show a clear tendency to increase.

\section{Usefulness of the Solutions Beyond $hh$ Discovery}
\label{sec:6}

 Beyond feature engineering for better data representation and separability, the neutrino solutions can be used to reconstruct interesting variables that could be lost otherwise. In fact, there are supervised, weakly supervised, and unsupervised ways to reconstruct a resonance that decays into missing energy using machine learning~\cite{Franceschini:2022vck, Alves:2022gnw, Alves:2024sai}; however, the simplest solution is to construct suitable kinematic functions of the final state particles; the problem is recovering the momenta of the final state neutrinos as we did, for example.

\begin{figure}[t]
    \centering
    \includegraphics[width=0.475\linewidth]{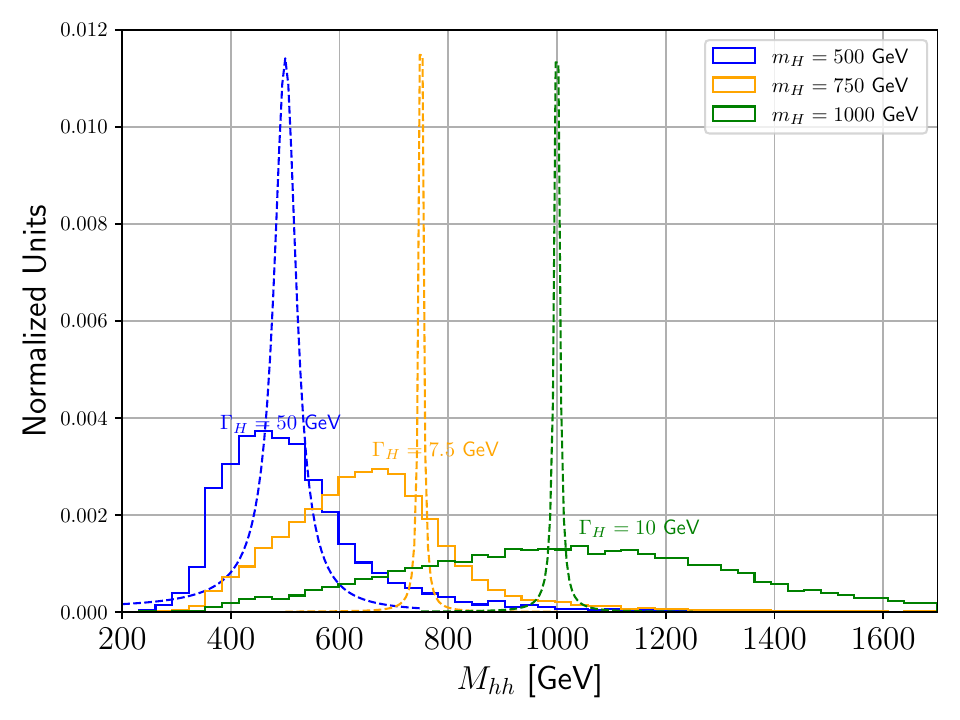}
    \includegraphics[width=0.425\linewidth]{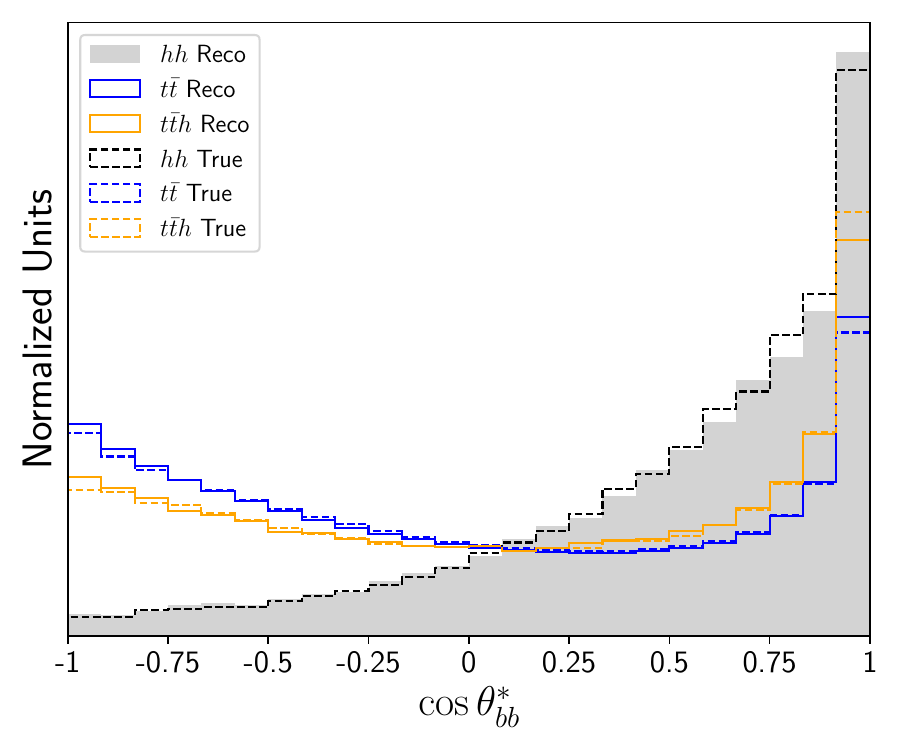}
\caption{In the left panel, we depict the invariant mass of the Higgs pair from the decay of a heavy Higgs boson, $H\to hh$, for masses of 500, 750, and 1000 GeV. The 500 GeV resonance has a 50 GeV width, while the others have 7.5 GeV and 10 GeV, respectively. The cosine of the angle between the bottom jets in their rest frame of the Higgs pair is shown in the right panel for the true and the reconstructed cases.}
    \label{fig:heavy}
\end{figure}

 As a first example, we display, in Figure~\ref{fig:heavy}, left panel, the invariant mass of Heavy Higgs bosons with masses and total widths of 500(50) GeV -- blue line, 750(7.5) GeV -- orange line, and 1000(10) GeV -- green line. The modes of the distributions are within a few percent of the true mass of Heavy Higgses, $mode(M_{hh})\sim m_H$, but the width of the distributions tends to be much larger than expected. The neutrino solution that compounds the invariant mass along with $\ell\ell+b\bar{b}$ is selected as $\nu_h$ if $\log H<5$ and as $\nu_t$ otherwise. 

 Another example is the reconstruction of the angle between the bottom jets in the reference frame where the Higgs bosons are back-to-back. Contrary to $t\bar{t}$, which is produced mostly at threshold, $hh$ is produced in a way where the Higgs bosons are somewhat boosted in that frame, making the bottom jets collimated. If the angle between the bottom jets is $\theta^*_{bb}$, then $d\sigma/d\cos\theta^*_{bb}$ is more concentrated towards 1, while $t\bar{t}$, and other backgrounds like $t\bar{t}h$, are more symmetric. This is what we observe in the right panel of Figure~\ref{fig:heavy}. This might be useful if one is interested in measuring coupling parameters that impact angular distributions, like axial and axial-vector couplings of Higgs bosons to fermions.
 As a final remark, there is a good agreement between the reconstructed (Reco) and true (True) parton-level neutrino solutions.

\section{Conclusions}
\label{sec:7}

 Once the existence of the Higgs boson, including its properties, has been firmly established, the LHC turns to a more challenging endeavor: the production of Higgs boson pairs. In order to study the scalar potential and measure the trilinear Higgs coupling, detecting the $hh$ production is necessary. Many channels have been proposed, and, currently, only combinations of many channels seem to lead to discovery prospects until 3000 fb$^{-1}$ is accumulated. 

 In this work, we use neutrino momenta solutions from the computation of two high-level variables, {\it Higgsness} and {\it Topness}, to engineer new variables which help to disentangle signals from backgrounds, mainly from huge $t\bar{t}$ and Drell-Yan production. We construct three new distinctive variables and optimize cuts in a variable space containing those new kinematic functions. We found a $3.7(3.6)\sigma$ statistical significance that favors the signal hypothesis with approximately 5(10) signal events and $S/B=6(2)$. These results are robust against systematics in the background yields.

 Beyond cut-and-count, we perform a multivariate analysis by training a BDT algorithm to separate the event classes. Instead of just placing cuts on the BDT output scores, we took advantage of the shapes of the distributions with a profile likelihood analysis. This time, we found that, keeping systematics at a low level of 5\%, the $5\sigma$ discovery is reachable.

 Beyond helping the discovery of the SM Higgs pair production, we also illustrate the usefulness of the neutrino solutions by reconstructing heavy scalar resonances and angular variables from SM Higgses decaying to $b\bar{b}\ell\ell\nu\nu$.

 \bigskip{}
	
\textbf{Acknowledgments}: This study was financed in part by Conselho Nacional de Desenvolvimento Científico e Tecnológico (CNPq), via the Grants No. 307317/2021-8 (A. A.), and in part by the Coordenação de Aperfeiçoamento de Pessoal de Nível Superior – Brasil (CAPES) – Finance Code 001 (D. S. V. G.). A. A. also acknowledges support from the FAPESP (No. 2021/01089-1) Grant. 

\bigskip{}

\appendix
\section{Critical points of {\it Topness}}
\label{apendiceA}

 In the construction of $RIV_t$ of Eq.~\eqref{eq:RIVt}, and $cM_{b_2\nu_{t_2}}$ of Eq.~\eqref{eq:cMbv}, we observe sharp peaks towards bins near zero, indicating that for some classes of events, mainly double Higgs production, the minimization of {\it Topness} leads to null solutions of the neutrino momenta, $p\nu,p_{\bar{\nu}}\approx 0$, so $\slashed M(\nu_t,\nu_t)\approx 0$ and the same for $M_{b_2\nu_{t_2}}$. 

 The logarithm of the {\it Topness} variable reads
 \begin{equation}
     \log T = \log\left[\sum_{i=1}^2\dfrac{(m^2_{\ell_i\nu_i}-m_W^2)^2}{s_W^4}+\sum_{i=1}^2\dfrac{(m^2_{\ell_i b_i\nu_i}-m_t^2)^2}{s_t^4}\right]\equiv \log J(p_\nu,p_{\bar{\nu}})
     \label{eq:topness2}
 \end{equation}
where $\nu_{1,2}$ can be a neutrino or anti-neutrino and just one of the branches $\chi_{12}$ or $\chi_{21}$ is active in the formula of Eq.~\eqref{eq:topness}. Notice that we set $s_W=s_t$ for top and $W$-boson resolutions, and we do not impose the measured missing transverse momentum constraint on the solutions, contrary to Ref.~\cite{Kim:2018cxf}.
\begin{figure}[t!]
    \centering
    \includegraphics[width=0.6\linewidth]{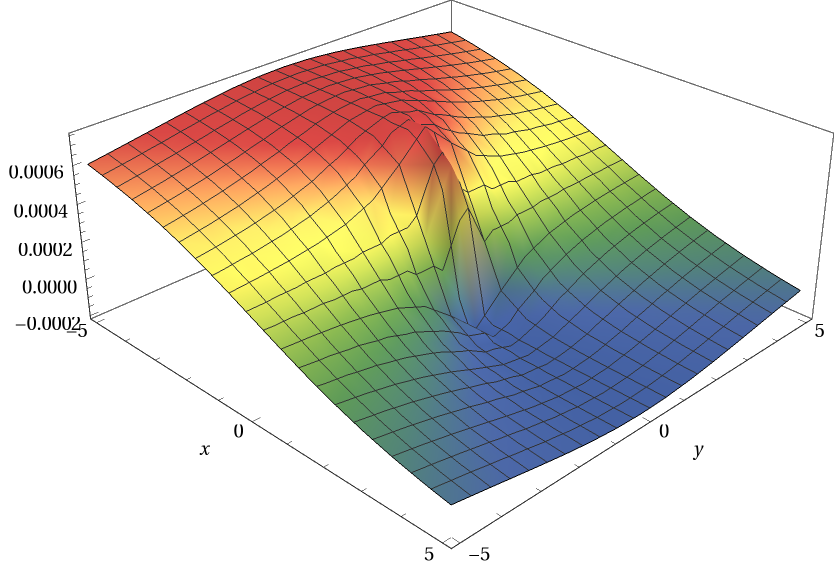}
\caption{The derivative of $\log T$ in respect do $x_\nu$ setting $z_\nu=0$. We clearly see the kink of the function at the origin where $p_\nu=0$, indicating a critical point.}
    \label{fig:dlogTdx}
\end{figure}
Let us concentrate on one of the neutrino components, if $p_\nu=(E_\nu,x_\nu,y_\nu,z_\nu)$, then the derivative of $\log T$ in respect to $x_\nu$, for example, is given by
\begin{equation}
    \dfrac{\partial \log T}{\partial x_\nu} = \dfrac{4}{J\ln 10}
    \left[\dfrac{m^2_{\ell\nu}-m_W^2}{s_W^4}\left(E_\ell\dfrac{x_\nu}{E_\nu}-p_{\ell x}\right)+\dfrac{m^2_{\ell b \nu}-m_t^2}{s_t^4}\left(E_{\ell b}\dfrac{x_\nu}{E_\nu}-p_{\ell b x}\right)\right]\; .
    \label{eq:diff}
\end{equation}

 For events not compatible with $t\bar{t}$ kinematics, $m^2_{\ell b\nu}\ne m_t^2$. In the case of $hh$ events, moreover, one of the $W$ bosons is produced off-shell, so $m_{\ell\nu}^2\ne m_W^2$. Also, notice that $J$ is pushed away from zero in this case. The critical points, therefore, could occur in two situations: (a) a very special case where the neutrino solution is collinear to both $\ell$ and $\ell+b$, or (b) $\vec{p_\nu},E_\nu \approx 0$ where $\log T$ is not differentiable, pushing the neutrino solutions toward null 4-momenta. In Figure~\ref{fig:dlogTdx}, we show the derivative of $\log T$ with $z_\nu=0$, where we clearly see the kink at the origin that corresponds to its critical point. Similar behavior is observed in the transverse mass variable, $m_T$~\cite{Cho:2007dh}.

\begin{figure}[t!]
    \centering
    \includegraphics[width=0.8\linewidth]{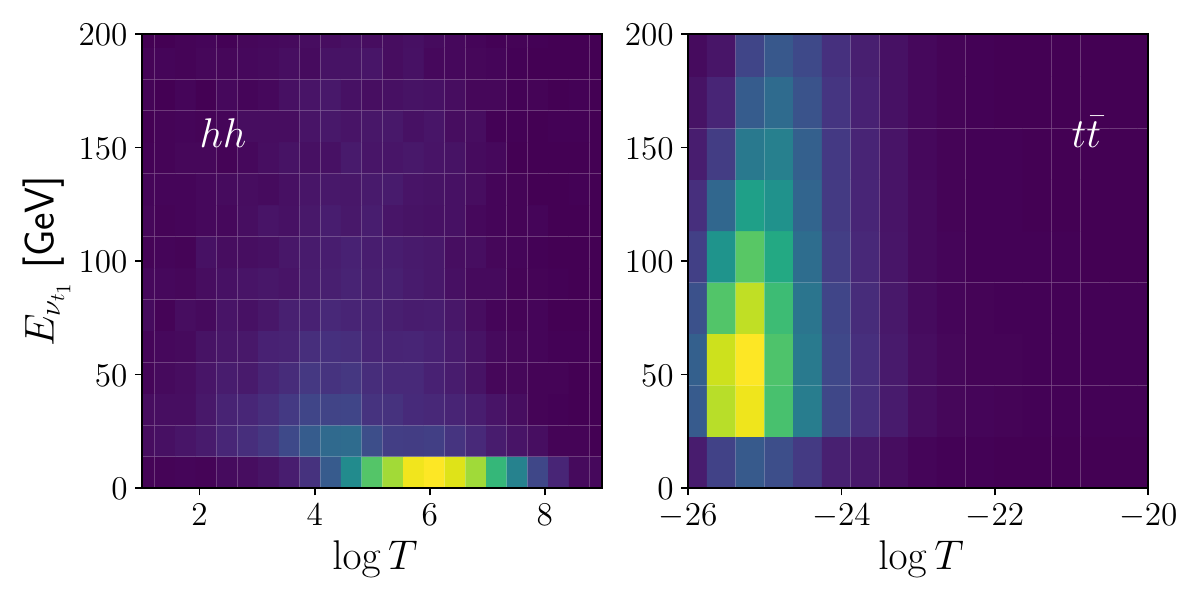}\\
    \includegraphics[width=0.8\linewidth]{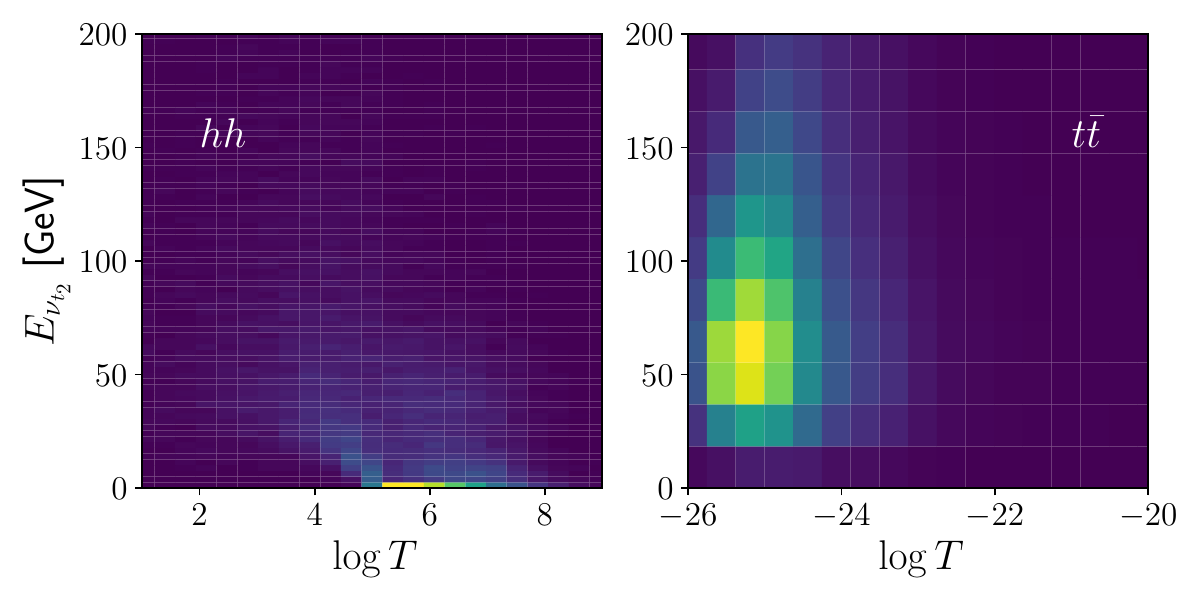}
\caption{The energy of neutrino solutions from {\it Topness} versus $log T$. In the left panels, we depict the $hh$ events, and in the right ones, the $t\bar{t}$ events.}
    \label{fig:evt-logT}
\end{figure}
 For $t\bar{t}$ events, though, the mass constraints that multiply the parentheses in Eq.~\eqref{eq:diff} are satisfied and the neutrino solutions do not need to vanish. As a consequence, invariant masses involving $\nu_t$ neutrinos will be hard for events with low {\it Topness} and soft for those with high {\it Topness} such as $hh$ events. We confirm these expectations by taking a look at the Figure~\ref{fig:evt-logT}, where we plot the energy of the two neutrino solutions from the minimization of {\it Topness}. As we see, $hh$ events, which have large $\log T$, present small neutrino energies, while $t\bar{t}$ events, which have much smaller $\log T$, have much larger energies.

\section{Momentum components of heavy particles decaying to neutrinos}
\label{apendiceB}

\begin{figure}[t]
    \centering
    \includegraphics[width=1.1\linewidth]{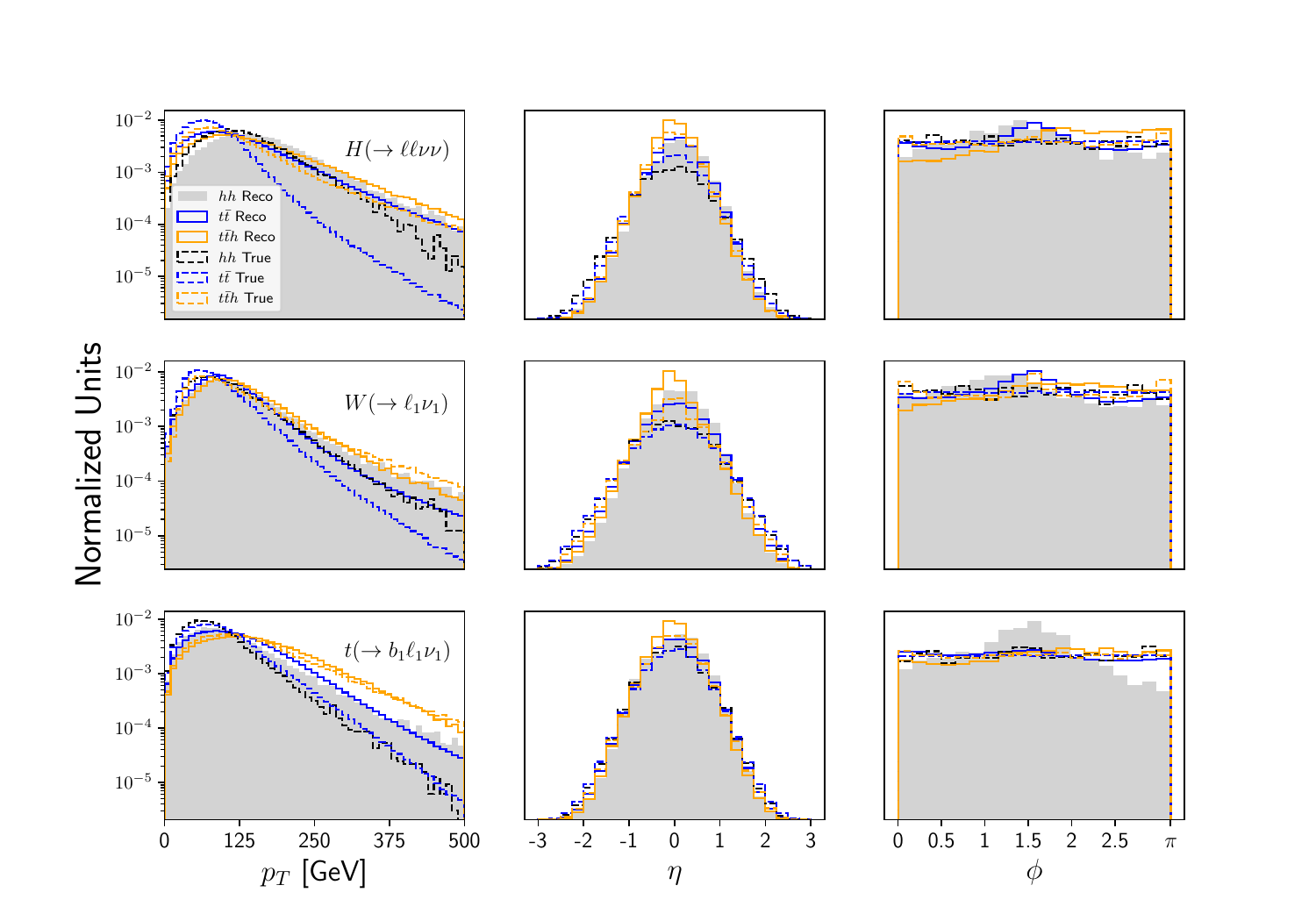}
\caption{In the left, middle, and right columns, we display the transverse momentum, $\eta$, and $\phi$, respectively, of $h\to \ell\ell\nu\nu$, the harder $W\to\ell\nu$ boson, and the harder top quark $t\to b\ell\nu$ in the upper, middle, and lower rows, respectively.}
    \label{fig:heavy2}
\end{figure}
In this appendix, we show the momentum components $(p_T,\eta,\phi)$ of the leptonic Higgs boson, $H\to \ell\ell\nu\nu$, the harder $W\to\ell\nu$ boson, and the harder top quark $t\to b\ell\nu$ in the upper, middle, and lower plots of Figure~\ref{fig:heavy2}. For the transverse momentum, the best agreement between true and reconstructed variables is observed for $t\bar{t}h$ events, followed by $hh$, and $t\bar{t}$. The same can be said about the rapidity, $\eta$. In terms of $\phi$, we observe the opposite behavior, where the best agreement occurs for $t\bar{t}$, and the worst is $t\bar{t}h$. Notice that $p_T$ is in log scale, while the other two variables are displayed in linear scale.

Interestingly, the $t\bar{t}h$ events, which mix the characteristics of both $hh$ and $t\bar{t}$ processes, present the best overall agreement between true and reconstructed neutrino momenta, which are reflected in these distributions. This opens up the possibility to study the chiral coupling of top quarks and the Higgs boson in $t\bar{t}h$ events with angular distributions in totally leptonic channels.


 %
%

\bibliography{myrefs}

\end{document}